\documentclass[amsmath,amssymb,aps,pre,floatfix,twocolumn,superscriptaddress]{revtex4}

\usepackage{changepage}
\usepackage{mathtools}
\usepackage{floatrow}
\usepackage{graphicx}
\usepackage{dcolumn}
\usepackage{bm}
\usepackage{tikz}
\usetikzlibrary{shapes, arrows, calc, positioning,matrix}
\usepackage[export]{adjustbox}
\usepackage{bbold}
\usepackage{graphicx}
\usepackage{hyperref}
\usepackage[font=small,justification=RaggedRight,labelfont=bf]{caption}
\usepackage{enumitem}
\usepackage[utf8]{inputenc}
\usepackage[T1]{fontenc}
\usepackage{soul,xcolor}
\usepackage{float}
\usepackage[label font=bf,labelformat=simple]{subfig}
\usepackage{multirow}

\makeatletter
\newcommand{\subalign}[1]{%
  \vcenter{%
    \Let@ \restore@math@cr \default@tag
    \baselineskip\fontdimen10 \scriptfont\tw@
    \advance\baselineskip\fontdimen12 \scriptfont\tw@
    \lineskip\thr@@\fontdimen8 \scriptfont\thr@@
    \lineskiplimit\lineskip
    \ialign{\hfil$\m@th\scriptstyle##$&$\m@th\scriptstyle{}##$\crcr
      #1\crcr
    }%
  }
}
\makeatother

\makeatletter
\newcommand{\shorteq}{%
  \settowidth{\@tempdima}{-}
  \resizebox{\@tempdima}{\height}{=}%
}
\makeatother

\begin{document}

\title{Correspondence between temporal correlations in time series, inverse problems, and the Spherical Model}


\author{Riccardo Marcaccioli}
\email{riccardo.marcaccioli.16@ucl.ac.uk}
\affiliation{Department of Computer Science, University College London, 66-72 Gower Street, London WC1E 6EA, UK}

\author{Giacomo Livan}
\email{g.livan@ucl.ac.uk}
\affiliation{Department of Computer Science, University College London, 66-72 Gower Street, London WC1E 6EA, UK}
\affiliation{Systemic Risk Centre, London School of Economics and Political Sciences, Houghton Street, London WC2A 2AE, UK}


\begin{abstract}
In this paper we employ methods from Statistical Mechanics to model temporal correlations in time series. We put forward a methodology based on the Maximum Entropy principle to generate ensembles of time series constrained to preserve part of the temporal structure of an empirical time series of interest. We show that a constraint on the lag-one autocorrelation can be fully handled analytically, and corresponds to the well known Spherical Model of a ferromagnet. We then extend such a model to include constraints on more complex temporal correlations by means of perturbation theory, showing that this leads to substantial improvements in capturing the lag-one autocorrelation in the variance. We apply our approach on synthetic data, and illustrate how it can be used to formulate expectations on the future values of a data generating process.
\end{abstract}

\maketitle

\section{Introduction}

During last two decades, multidisciplinary applications of physics - ranging from economics~\cite{econ_1,econ_2} and finance~\cite{finance_1,finance_2} to sociology~\cite{socio_1,socio_2}, biology~\cite{bio_1,bio_2} and linguistics \cite{language_1,language_2} - have witnessed an increasing attention from the physics community~\cite{inter_phys_1,inter_phys_2,inter_phys_3,inter_phys_4}. Indeed, physicists have contributed to the development of methodologies that are of crucial importance to such disciplines, such as, e.g., network modelling and analysis \cite{networks_newman_book,polya}, game theory \cite{game_theory_1,game_theory_2} and time series analysis \cite{time_series_1,time_series_2}. 

Arguably, one of the main driving forces of such an interest is Statistical Mechanics, which provides a unified and coherent framework based on first principles to model large interacting systems even outside the realm of Physics. In particular - as originally suggested by Jaynes \cite{jaynes_1,jaynes_2} - the Maximum Entropy principle has been used as a flexible tool to build unbiased statistical models in a vast range of different disciplines \cite{max_ent_real_net,max_ent_geo,max_ent_brain}. 

However, in most of such applications, the Maximum Entropy principle is used in the opposite way with respect to its classical use in Statistical Mechanics, where the goal is usually to compute observable macroscopic quantities (such as correlations in an Ising model) from the unobservable microscopic laws ruling the interactions between the components of a system \cite{zecchina_rev}. The opposite problem is that of inferring the parameters of an interacting system (e.g., the coupling constants and fields in an Ising model) from snapshots of its microscopic configurations. This is referred to as the \textquotedblleft inverse problem\textquotedblright. In Physics, it has received considerable attention especially when applied to fully connected Ising models \cite{zecchina_rev,ising_1}. Outside Physics, instead, it has provided a theoretical basis for some of the aforementioned interdisciplinary applications, due to the increased accessibility of the ``microscopic configurations'' of many non-physical systems (e.g., financial markets, social networks, neuron firing patterns, etc.).

Jaynes proposed an alternative principle when dealing with models of time-evolving systems, typically non stationary or out of equilibrium ones. Known as the Maximum Caliber principle \cite{max_cal_jaynes}, its goal is to determine an unbiased distribution over all possibles paths of a system by maximising the system's path entropy while preserving some desired constraints on its trajectories. Researchers have used the Maximum Caliber principle in a wide rage of different applications \cite{max_cal_1,max_cal_2}, the majority of which have been devoted to determining the transition rates of Markov models \cite{markov_max_cal_1,markov_max_cal_2} for systems evolving in continuous time between a fixed set of states. 

In the case of systems evolving in discrete time, the Maximum Caliber principle can be shown to coincide with the Maximum Entropy principle by mapping time as the spatial dimension of a lattice whose sites are occupied by events \cite{max_cal_equivalence,old_p}. In practice, this mapping effectively corresponds to \emph{time series}, as even most systems evolving in continuous time are sampled at discrete times. 

In Ref. \cite{old_p}, we have shown how the Maximum Caliber / Entropy formulation can be used to generate ensembles of multivariate time series in discrete time constrained to preserve - on average - some empirically observed distributional properties of a multivariate system (such as, e.g., higher order moments and seasonalities). One of the main challenges of the multivariate case presented in \cite{old_p} is that of explicitly accounting for correlations, which can only be captured indirectly via other constraints.

In the present work, we partially overcome such limitations by tackling the problem of explicitly accounting for temporal correlations in the case of univariate systems. Modeling the temporal correlations of statistical systems is a notoriously challenging task. The most frequently used tools are autoregressive models belonging to the ARCH-GARCH family \cite{AR,AR2}, or stochastic processes such as the Ornstein-Uhlenbeck model \cite{O-U}. These models - and those inspired by them - have enjoyed great success in a variety of applications where modeling time correlations can be crucial, such as, e.g., in Economics or Finance \cite{cc1,cc2}. Here, instead, we adopt a data-driven perspective - grounded in the Maximum Entropy principle - in order to capture the time correlations of a system without the need to explicitly model its time evolution.

First, we will briefly introduce the general methodology in its full mathematical form (Section \ref{sec:maxent}). Then, we will apply it to generate ensembles of time series designed to preserve on average correlations between first moments as measured in an empirical time series of interest. In order to do so, we will leverage the Spherical Model \cite{spherical_model_original,spherical_model_2}, and we will show that it naturally corresponds to autoregressive processes \cite{AR} (Section \ref{sec:first_ham}). After that, we will proceed to account for higher order temporal correlations. We will do so by solving a more complex model by expanding on the Spherical Model by means of perturbation theory (Section \ref{sec:complex_ham}). In the former case, the analytical knowledge of the Spherical Model's partition function makes the calibration of the proposed approach extremely simple, whereas in the latter case we will show how the Plefka expansion \cite{plefka} - a technique commonly used for the inverse Ising problem - can be applied to find an approximate solution.  

\section{Maximum Entropy framework for time series data}
\label{sec:maxent}

Let $\mathcal{X}$ be the set of all real-valued time series of length $T$, and let $\overline{X} \in \mathcal{X}$ be an empirical time series of interest, i.e., $\overline{x}_{t}$ stores the time $t$ value sampled from a variable under consideration. The goal of the methodology is to define an ensemble able to preserve - as ensemble averages - $L$ empirical measurements on $\overline{X}$. In other words, we want to find a probability density function $P(X)$ over $\mathcal{X}$, such that the expectation values $ \langle \mathcal{O}_\ell(X) \rangle = \sum_{X \in \mathcal{X}} \mathcal{O}_\ell(X) P(X) $ of a set of observables ($\ell = 1, \ldots, L$) coincide with their values measured in the given time series $\overline{O}_\ell = \mathcal{O}_\ell(\overline{X})$. In these terms, the problem is ill-defined, as $P(X)$ may be defined in an arbitrary number of ways. However, if we require $P(X)$ to also maximise the entropy $S(X) = \sum_{X \in \mathcal{X}} - P(X) \ln{(X)}$, computing $P(X)$ becomes a constrained maximization problem which can be uniquely solved by choosing:
\begin{equation*}
    P(X) = \frac{e^{- H(X)}}{Z} \ ,
\end{equation*}
 where $H(X) = \sum_\ell \beta_\ell \ \mathcal{O}_\ell(X)$ is the Hamiltonian of the ensemble, $\beta_\ell$ ($\ell = 1, \ldots, L$) are Lagrange multipliers introduced to enforce the constraints, and $Z =\sum_X e^{- H(X)}$ is the partition function of the ensemble, which verifies $ \langle \mathcal{O}_\ell(X) \rangle = - \partial \ln Z / \partial \beta_\ell \; , \forall \, \ell $. The existence and uniqueness of the Lagrange multipliers can be proved, and it can also be shown that they are equivalent to those that maximize the likelihood of drawing the time series $\overline{X}$ from the ensemble \cite{lagr_mult_coincides_maximum_like}.
 
The problem of determining $P(X)$ has therefore been solved. However, explicitly computing the Lagrange multipliers $\beta_\ell$ that maximise the likelihood of drawing the data from the ensemble without an analytical form for $Z$ can only be achieved by means of Boltzmann learning gradient-descent algorithms \cite{zecchina_rev}. These ultimately require an exhaustive phase space exploration through sequential Monte Carlo simulations, which quickly becomes computationally unfeasible for $T \gg 1$. Therefore, finding a closed form solution (even an approximate one) for $Z$ is the cardinal problem to be solved in order to fully define a working methodology. 

As a dummy example to illustrate how a specific ensemble can be computed, let us consider an empirical time series $\overline{X}_t$ of length $T$ and let us choose as constraints its sample mean $\overline{m} = \sum_{t=1}^T \overline{x}_t/T$ and mean square value $\overline{V} = \sum_{t=1}^T \overline{x}_t^2/T$. In order to compute the partition function $Z$, let us denote as $x_t$ the $t$-th element in $X$, and let us place each of such elements on a one dimensional lattice of length $T$. The constraints on the mean and mean square value lead to the following Hamiltonian:
\[
H = \sum_{t = 1}^{T} \left[ \lambda_1 x_t + \lambda_2 x_t^2 \right] .
\]
After having specified the constraints, what is left to do is to evaluate the partition function. In order to do that, we need to properly define the sum over the phase space $\mathcal{X}$ appearing in the definition of $Z$:
\[
\begin{aligned}
Z &= \sum_{X \in \mathcal{X}} e^{-H(X)} = \int_{-\infty}^{+\infty} \prod_{t=1}^{T} \,dx_{t} \; e^{-H(X)} = \\
&= \prod_{t=1}^{T} \int_{-\infty}^{+\infty} dx_{t} \;  e^{-\lambda_1 x_t - \lambda_2 x_t^2} = \left( \sqrt{\frac{\pi}{\lambda_2}}e^{\frac{\lambda_1^2}{4 \lambda_2}} \right)^T  ; \;\lambda_2>0 \ .
\end{aligned}
\]
Once the partition function is known, the Lagrange multipliers $\lambda_1$ and $\lambda_2$ can be found by solving the following system of coupled equations:
\[
\begin{aligned}
&\overline{m} = - \frac{1}{T} \frac{\partial \ln{Z}}{\partial \lambda_1} = - \; \frac{\lambda_1}{2 \lambda_2} \\
&\overline{V} = - \frac{1}{T} \frac{\partial \ln{Z}}{\partial \lambda_2} = \; \frac{\lambda_1^2+ 2 \lambda_2}{4 \lambda_2^2} \ ,
\end{aligned}
\]
which leads to the following probability density function for the ensemble:
\[
P(X) = \left ( \frac{1}{2 \pi ( \overline{V}- \overline{m}^2)} \right )^{T/2} \prod_{t=1}^{T} \; e^{- \frac{(x_t-\overline{m})^2}{2 (\overline{V} - \overline{m}^2)}}  \ ; \quad V>m^2 \ ,
\]
which is the factorized probability density function of $T$ independent Gaussian random variables with mean $\overline{m}$ and variance $(\overline{V} - \overline{m})^2$.

\section{The first Hamiltonian}
\label{sec:first_ham}

We shall now apply the framework introduced in the previous section to a more complex set of constraints, namely the sample mean ($\overline{m}$), mean square value ($\overline{V}$) and temporal correlation at lag-one $\overline{C}_1 = \sum_{t=1}^T \overline{x}_t \overline{x}_{t+1}$ (notice that the following steps generalize to a generic temporal correlation $\overline{C}_\tau = \sum_{t=1}^T \overline{x}_t \overline{x}_{t+\tau}$). 

Let us place the data points on a one-dimensional temporal lattice, whose sites $t = 1, \ldots, T$ correspond to the events of a time series of interest $\overline{x}_1, \ldots,  \overline{x}_T$. After doing that, the specified set of constraints leads to the following Hamiltonian:
\begin{equation}\label{eq: hamiltonian spherical model}
    H = \sum_{t=1}^T \left[ \lambda_1 x_t + \lambda_2 x_t^{2} + \lambda_3 x_t x_{t+1}  \right] \; ,
\end{equation}
where we are assuming spherical boundary conditions $x_{T+1} = x_1$. The Hamiltonian in Eq.~\eqref{eq: hamiltonian spherical model} is that of the Spherical Model~\cite{spherical_model_original}, a well-known model in Statistical Mechanics. 

Having specified the Hamiltonian, the task now becomes finding the partition function $Z$, which reads:
\begin{equation}\label{eq: gauss integral}
\begin{aligned}
Z &= \int_{-\infty}^{+\infty} \prod_{t=1}^{T} \,dx_{t} \; e^{-\lambda_1 x_t - \lambda_2 x_t^{2} - \lambda_3 x_t x_{t+1}} = \\
&= \int d^T x \; e^{- x^\mathrm{T} A x + B^\mathrm{T} x} = \sqrt{\frac{\pi^T}{\det{A}}}\; e^{\frac{B^\mathrm{T} A^{-1} B}{4}} \; ,
\end{aligned}
\end{equation}
where we have introduced the following vector notation:
\begin{equation*}
\small{
B^\mathrm{T} = -\lambda_1 \begin{pmatrix}
1\\ 
\vdots\\ 
1
\end{pmatrix}, \ 
A = \begin{pmatrix}
\lambda_2 & \lambda_3 & 0 & \cdots & \cdots & \cdots & \cdots & \lambda_3\\
\lambda_3 & \lambda_2 & \lambda_3 & 0 & \cdots & \cdots & \cdots & 0\\
0 & \lambda_3 & \lambda_2 & \lambda_3 & \ddots & & & \vdots\\
\vdots & 0 & \ddots & \ddots & \ddots & \ddots & & \vdots\\
\vdots & & \ddots & \ddots & \ddots & \ddots & 0 & \vdots\\
\vdots & & & \ddots & \lambda_3 & \lambda_2 & \lambda_3 & 0\\
\vdots & & & & 0 & \lambda_3 & \lambda_2 & \lambda_3\\
\lambda_3 & 0 & \cdots  & \cdots & \cdots & 0 & \lambda_3 & \lambda_2\\
\end{pmatrix}} \ .
\end{equation*}
Using the fact that $A$ is a special case of a real symmetric circulant matrix (whose spectral properties are generally known \cite{circulant_matrices}), one can show that its eigenvalues are $\Lambda_t = \lambda_2 + \lambda_3 \cos{ \frac{2 \pi}{T} (t-1) }$ ($t = 1,\ldots,T$). These can be used to highlight - in the limit $T \gg 1$ - the explicit dependency of $Z$ from the Lagrange multipliers. Expanding each term appearing in Eq.~\eqref{eq: gauss integral}, we have:
\begin{equation} \label{eq:calculation parts of Z}
\begin{aligned}
&\det{A} = \prod_{t=1}^T \Lambda_t =  e^{ \sum_{t=1}^T \ln{\left[\lambda_2 + \lambda_3 \cos{ \frac{2 \pi}{T} (t-1) } \right]}} \\
& \qquad \; \approx e^{ \frac{T}{2 \pi} \int_{0}^{2 \pi} d\omega \, \ln{\left[\lambda_2 + \lambda_3 \cos{\omega} \right]}} = e^{ T \ln{ \frac{\lambda_2+\sqrt{\lambda_2^2-\lambda_3^2}}{2} }} \\
& \qquad \;\;= \left(\frac{\lambda_2+\sqrt{\lambda_2^2-\lambda_3^2}}{2} \right)^T \ ; \\ 
& B^\mathrm{T} A^{-1} B = \lambda_1^2 \; b^\mathrm{T} A^{-1} b = \lambda_1^2 \; b^\mathrm{T} \frac{1}{\Lambda_1} b = \; T \; \frac{\lambda_1^2}{\lambda_2+\lambda_2} \ ,
\end{aligned}
\end{equation}
where we have used the fact that $b = (1,\ldots,1)$ is the eigenvector of $A$ (and therefore of $A^{-1}$) associated to $\Lambda_1$. Plugging the above expressions into Eq.~\eqref{eq: gauss integral}, we obtain the ensemble's partition function, which reads:
\begin{equation}\label{eq: Z spherical}
    Z = \left( \frac{2 \pi}{\lambda_2+\sqrt{\lambda_2^2-\lambda_3^2}} \right)^{\frac{T}{2}} e^{T \, \frac{\lambda_1^2}{4 ( \lambda_2+\lambda_2 )}} \; .
\end{equation}
From Eq.~\eqref{eq: Z spherical} we can derive the system of equations for the Lagrange multipliers:
\begin{equation}\label{eq: partial der Z sph}
\begin{aligned}
\frac{\overline{m}}{T} &= - \frac{\lambda_1}{2 ( \lambda_2+\lambda_3 )} \\
\frac{\overline{V}}{T} &= \frac{\lambda_1^2}{4 ( \lambda_2+\lambda_3 )^2} + \frac{1}{2 \sqrt{\lambda_2^2-\lambda_3^2}} \\
\frac{\overline{C}_1}{T} &= \frac{\lambda_1^2}{4 ( \lambda_2+\lambda_3 )^2} + \frac{\lambda_3}{2 (\lambda_3^2 - \lambda_2^2 + \lambda_2 \sqrt{\lambda_2^2-\lambda_3^2})} \ .
\end{aligned}
\end{equation}
The above equations can be easily solved analytically. Their expressions are not particularly instructive, so we omit them for easiness of exposition. Once the system in Eq.~\eqref{eq: partial der Z sph} has been solved the ensemble is fully defined, and instances can be drawn from it with standard Monte Carlo methods~\cite{monte_carlo}. 

\floatsetup[figure]{style=plain,subcapbesideposition=top}
\begin{figure*} 
  \centering
  \includegraphics[width=1\linewidth]{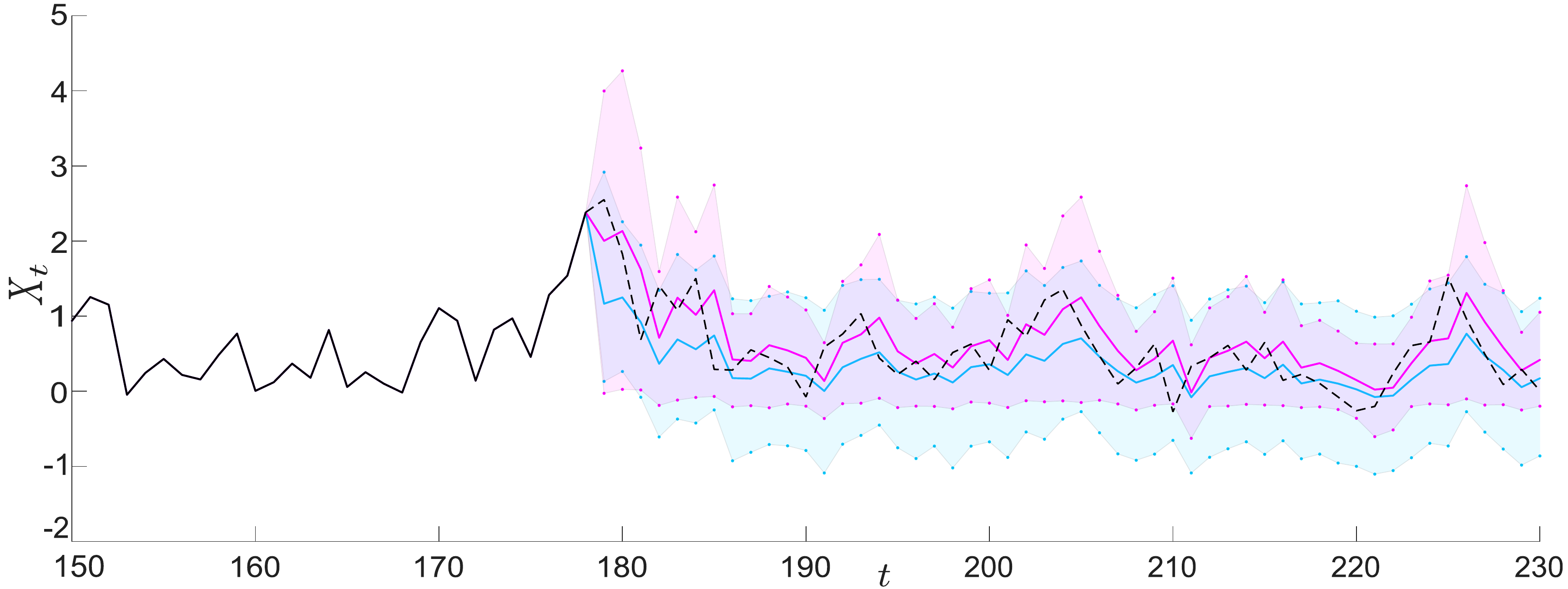}
  
  \caption{\textbf{Reconstruction of a known data generating process}. Black lines in the Figure correspond to data generated synthetically from the autoregressive model $Y_{t+1} = \xi_t^{(0,1.5)} Y_t + \xi_t^{(-0.3,0.7)}$. The solid black line corresponds to data up to time $T=180$, which are used to compute the initial values of the ensemble's Lagrange multipliers appearing in Eq.~\eqref{eq: Z spherical}, while the black dashed line corresponds to the evolution of the process beyond time $T$. The blue (light grey) solid line and shaded region denote, respectively, ``out of sample'' next-step expectations for times $t > T$ based on the ensemble, with Lagrange multipliers updated in ``real time'' based on new data points. The purple (dark grey) solid line and shaded region correspond, respectively, to the mean and $99\%$ confidence interval computed over a sample of $10^6$ trajectories of the process $X_t$ generated as one-step increments starting - at all times - from the values represented by the dashed black line.}
  \label{fig: pred sph}
\end{figure*}

Figure~\ref{fig: pred sph} shows an application of the ensemble aimed at reconstructing a known data generating process. Black lines correspond to data generated synthetically from an autoregressive model defined as follows: $Y_{t+1} = \, \xi_t^{(a_1,a_2)} Y_t + \xi_t^{(a_3,a_4)}$, where $\xi_t^{(a,b)}$ is a random number drawn at time $t$ from a uniform distribution in the interval $[a,b]$. The solid black line corresponds to the final $30$ points of an initial time series of length $T = 180$, which we use to compute the Lagrange multipliers appearing in Eq.~\eqref{eq: Z spherical} for the first time. The black dashed line corresponds to the continuation of such time series beyond time $T$, which we use both to update the Lagrange multipliers in ``real time'', and to test the agreement between the scenarios generated by the ensemble with respect to new data points. The blue (light grey) solid line and shaded region correspond to ``out of sample'' next-step expectations for times $t > T$ (i.e., obtained by recomputing the ensemble's Lagrange multipliers for all times $t \geq T+1$), denoting, respectively, the average value and $99\%$ confidence interval computed from the ensemble via Monte Carlo simulations. The purple (dark grey) solid line and shaded region instead capture the ``true'' next-step evolution of the system. They correspond, respectively, to the mean and $99\%$ confidence interval computed over a sample of $10^6$ trajectories of the aforementioned autoregressive model generated as one-step increments starting - at all times - from the values represented by the dashed black line. As it can be seen from a qualitative inspection of Figure~\ref{fig: pred sph}, the ensemble reproduces rather faithfully the average time evolution of the underlying data generating process. There are, however, some visible deviations between the two confidence intervals shown in Figure~\ref{fig: pred sph}. These are due to the fact that the data generating process has non trivial time correlations in its higher order moments, which are not captured by the ensemble. These will be captured by the model introduced in the next Section, where we will also perform a more rigorous statistical assessment of the model's ability to reconstruct a data generating process.

\section{A more complex Hamiltonian}
\label{sec:complex_ham}

We now proceed to investigate a more complex ensemble encoding additional constraints. We consider the following Hamiltonian:
\begin{equation}\label{eq: hamiltonian quartic}
    H = \sum_{t=1}^T \left[ \lambda_1 x_t + \lambda_2 x_t^{2} + \lambda_3 x_i x_{t+1} + \lambda_4 x_t^{2} x_{t+1}^2 + \lambda_5 x_t^{4}  \right] \ ,
\end{equation}
which enforces the constraints already considered in the Hamiltonian of Eq.~\eqref{eq: hamiltonian spherical model}, plus additional constraints on the sample mean fourth power ($\sum_{t=1}^T \overline{x}_t^4$) and on the time correlations at lag-one between squared values ($\sum_{t=1}^T \overline{x}_t^2 \overline{x}_{t+1}^2$). Such constraints - coupled with the ones mentioned previously - effectively amount to constraining, respectively, the ensemble average on the kurtosis and on the variance autocorrelation at lag-one.

Similarly to Eq.~\eqref{eq: gauss integral}, the partition function resulting from Eq.~\eqref{eq: hamiltonian quartic} reads
\begin{equation}\label{eq: Z quartic}
    Z = \int_{-\infty}^{+\infty} \prod_{t=1}^{T} \,dx_{i} \; e^{-\lambda_1 x_t - \lambda_2 x_t^{2} - \lambda_3 x_t x_{t+1} + \lambda_4 x_t^{2} x_{t+1}^2 + \lambda_5 x_t^{4}} \ .
\end{equation}
Integrals similar to the one above appear in $\lambda \phi^4$ lattice field theories, and are known for not being solvable analytically. However, such calculations are commonly tackled by using resummation techniques or perturbation theory \cite{lattice_qft}. Following this line of research, we will make use of the Plefka expansion - a perturbation method widely used in the inverse Ising problem -  in order to find approximate estimates of the true Lagrange multipliers.

In standard perturbation theory, the true Hamiltonian $H$ of a system is written as a sum of an unperturbed part $H_0$ and a perturbation $H_p$, i.e., $H = H_0+H_p$. Using this notation, the partition function of the system becomes:
\begin{equation}\label{eq: pert gen}
\begin{aligned}
    Z &= \sum_{\mathcal{X}} e^{-(H_0+H_p)} = Z_0 \sum_{\mathcal{X}} \frac{e^{-H_0}}{Z_0}e^{-H_p} \\
    &= Z_0 \langle e^{-H_p} \rangle_0 = Z_0 \sum_k \frac{(-1)^k}{k !} \langle H_p^k \rangle_0 \ ,
\end{aligned}
\end{equation}
where $Z_0$ is the partition function of the unperturbed system ($Z_0 = \sum_{\mathcal{X}} e^{-H_0}$), and $\langle \cdots \rangle_0$ is the average over the ensemble defined by $Z_0$. Equation~\eqref{eq: pert gen} is exact. However, in order to make it usable in practise, one needs to truncate the power series expansion (which becomes a power series expansion in the Lagrange multipliers appearing in the definition of $H_p$) to a certain order $k$. Of course, if one is lucky enough to find a recursion for $\langle H_p^k \rangle_0$ and to sum the resulting series, one can in principle compute the true partition function $Z$. 

The Plefka expansion follows a very similar procedure to the one just described. It starts from the Hamiltonian of the system written as $H = H_0 + \lambda H_p$, where $\lambda$ is a constant that serves to distinguish different perturbation orders which will be ultimately set to one. Instead of expanding the partition function $Z$, the Plefka expansion considers the free energy of the system:
\begin{equation}
    F = - \ln Z = - \ln Z_0 - \ln \frac{Z}{Z_0} = F_0 + F_p \ ,
\end{equation}
where $F_0$ is the free energy of the unperturbed ensemble and $F_p = - \ln \frac{Z}{Z_0}$. We can now expand $F_p$ as a power series in $\lambda$:
\begin{equation}\label{eq: boh 2}
    F_p = - \lambda f_1 + \frac{\lambda^2}{2} f_2 - \frac{\lambda^3}{3 !} f_3 + \cdots 
\end{equation}
where we used the fact that if $\lambda=0$ then $F = F_0$. Substituting into $e^{-F_p} = Z/Z_0$, we obtain:
\begin{equation}\label{eq: boh}
    \frac{Z}{Z_0} = 1 - \lambda f_1 + \frac{\lambda^2}{2} (f_2 + f_1^2) - \frac{\lambda^3}{3 !} ( f_3 + f_1^3 + 3 f_2 f_1 ) + \cdots
\end{equation}
Comparing Eq.~\eqref{eq: boh} with the direct power series expansion $Z/Z_0 = \sum_k (-\lambda)^k \langle H_p^k \rangle_0 / k! $, we obtain an explicit expression for every term of the expansion in Eq.~\eqref{eq: boh 2}:
\begin{equation}\label{eq: plefka exp}
    \begin{aligned}
    f_1 &= \langle H_p \rangle_0 \\
    f_2 &= \langle H_p^2 \rangle_0 - f_1^2 \\
    f_3 &= \langle H_p^3 \rangle_0 - f_1^3 - 3 f_1 f_2 \ .
    \end{aligned}
\end{equation}
As it can seen from Eq.~\eqref{eq: plefka exp}, the expansion of the free energy $F$ is effectively an expansion around the cumulants of the unperturbed ensemble. A similar idea was developed (8 years earlier than Plefka) by Bogolyubov \textit{et al.} \cite{plefka_bogolyubov} for the ferromagnetic Ising model.

Let us now perform a second order Plefka expansion in the case of the Hamiltonian in Eq.~\eqref{eq: hamiltonian quartic}. We will consider the Spherical Model~\eqref{eq: Z spherical} as the unperturbed ensemble $Z_0$, with the perturbation given by $H_p = \sum_t \left[ \lambda_4 x_t^2 x_{t+1}^2 + \lambda_5 x_t^2 \right]$. As a result, the second order approximated free energy reads
\begin{equation} \label{eq: plefka exp sph}
    \begin{aligned}
    F \approx F_0 &- \sum_t \left[ \lambda_5 \langle x_t^4 \rangle_0 + \lambda_4 \langle x_t^2 x_{t+1}^2 \rangle_0 \right] \\
    &+ \frac{1}{2} \sum_{t,t^\prime} \left[ \lambda_5^2 \langle x_t^4 x_{t^\prime}^4 \rangle_0 + \lambda_4^2 \langle x_t^2 x_{t+1}^2 x_{t^\prime}^2 x_{t^\prime+1}^2 \rangle_0 \right. \\
    & \qquad \qquad \left. + 2 \lambda_5 \lambda_4 \langle x_t^4 x_{t^\prime}^2 x^2_{t^\prime+1} \rangle_0 \right] \\
    & - \frac{1}{2} \sum_{t} \left[ \lambda_5 \langle x_t^4 \rangle_0 + \lambda_4 \langle x_t^2 x_{t+1}^2 \rangle_0 \right]^2 \ ,
    \end{aligned}
\end{equation}
where the expansion above has introduced a second time index $t^\prime$. In the following, we shall make use of this in order to introduce distances between sites $t - t^\prime$, which correspond to temporal distances between events in the original time series.

We now proceed to evaluate the expectation values in Eq.~\eqref{eq: plefka exp sph} around the ensemble defined by Eq.~\eqref{eq: Z spherical}. In order to do that, we need to  apply Isserlis’ Theorem~\cite{ISSERLIS}, a result which is also largely employed in quantum field theory under the name of Wick's Theorem \cite{GIANCARLO_WICK}.

For easiness of exposition, let us redefine some quantities appearing in Eq~\eqref{eq: partial der Z sph} as follows:
\begin{equation} \label{eq: m,s0,s1,stt'}
\begin{aligned}
    m &= - \frac{\lambda_1}{2 ( \lambda_2+\lambda_3 )} = - \frac{\partial}{\partial \lambda_1} \ln Z_0 \\
    s_0 &= \frac{1}{2 \sqrt{\lambda_2^2-\lambda_3^2}} = - \frac{\partial}{\partial \lambda_2} \ln Z_0 \bigg|_{\lambda_1=0} \\
    s_1 &=  \frac{\lambda_3}{2 (\lambda_3^2 - \lambda_2^2 + \lambda_2 \sqrt{\lambda_2^2-\lambda_3^2})} = - \frac{\partial}{\partial \lambda_3} \ln Z_0 \bigg|_{\lambda_1=0} \\
    s_{t t^\prime} &= \langle x_t x_{t^\prime} \rangle_0 |_{\lambda_1=0} \ ,
    \end{aligned}
\end{equation}
where $s_0 = s_{tt}$ and $s_1 = s_{t,t+1}$, $\forall t$.

We can now proceed to calculate the expectation values appearing in Eq.~\eqref{eq: plefka exp sph}. These read
\begin{equation} \label{eq: sph moments}
\small{
    \begin{aligned}
        &\langle x_t^4 \rangle_0 = m^4 + 6 m^2 s_0 + 3 s_0^2 \\
        &\langle x_t^2 x_{t+1}^2 \rangle_0 = (m^2 + s_0)^2 + 4 m^2 s_1 + 2 s_1^2 \\
        &\langle x_t^4 x_{t^\prime}^4 \rangle_0 = (m^4 + 6 m^2 s_0 + 3 s_0^2)^2 + 16 (m^3 + 3 m s_0)^2 s_{t t^\prime} \\
        & \qquad \qquad \;\; + 72 (m^2 + s_0)^2 s_{t t^\prime}^2 + 96 m^2 s_{t t^\prime}^3 + 24 s_{t t^\prime}^4 \\
        &\langle x_t^4 x_{t^\prime}^2 x_{t^\prime+1}^2 \rangle_0 = f(m,s_0,s_1,s_{t t^\prime},s_{t,t^\prime+1}) \\
        &\langle x_{t}^2 x_{t+1}^2 x_{t^\prime}^2 x_{t^\prime+1}^2 \rangle_0 = g(m,s_0,s_1,s_{t t^\prime},s_{t, t^\prime+1},s_{t+1, t^\prime}) \ ,
    \end{aligned}}
\end{equation}
where $f$ and $g$ are polynomial functions of their variables and are specified in the Appendix. 

As one can see from Eqs.~\eqref{eq: plefka exp sph} and~\eqref{eq: sph moments}, the second order approximation contains the covariances of the unperturbed Hamiltonian at all possible ranges, i.e., not just at lag-one. As a result, we need to find an explicit form for $s_{t t^\prime}$ in order to move forward. Following the steps that lead to the solution of the Gaussian integral in Eq.~\eqref{eq: gauss integral}, we have:
\begin{equation}\label{eq: sph autocorr}
\begin{aligned}
     s_{t t^\prime} &= \langle x_t x_{t^\prime} \rangle_0 |_{\lambda_1=0} = \langle \sum_{s} V_{t s} y_s \sum_{k} V_{t^\prime k} y_k  \rangle_0 \bigg |_{\lambda_1 = 0}  \\ 
     &= \sum_{s,k} V_{t s} V_{t^\prime k} \langle y_s y_k \rangle_0 |_{\lambda_1=0} = \sum_s V_{t s} V_{t^\prime s} \langle y_s^2 \rangle_0 |_{\lambda_1 \text{\shorteq} 0} \\
     &= \sum_s \frac{1}{2 T} \frac{\cos{\left[ \frac{2 \pi}{T} (s-1) (t-t^\prime) \right]}}{\lambda_2 + \lambda_3 \cos{\left[ \frac{2 \pi}{T} (s-1) \right]}} \ ,
\end{aligned}
\end{equation}   
where $V_{t k} = \frac{1}{\sqrt{{T}}} ( \cos{\left[ \frac{2 \pi}{T} (t-1) (k-1) \right]} + \sin{\left[ \frac{2 \pi}{T} (t-1) (k-1) \right]} )$ is the $t$-th element of the $k$-th eigenvector of the matrix $A$ in Eq.~\eqref{eq: gauss integral}, and $y_k = \sum_{t} V_{tk} x_t$. The above expression can be then rewritten as
\begin{equation}
\begin{aligned}
s_{t t^\prime} &= \sum_s \frac{1}{2 T} \cos{\left[ \frac{2 \pi}{T} (s-1) R \right]} \; \times \\ &\qquad \times \; \int_0^{\infty} dz \, e^{-z \left( \lambda_2 + \lambda_3 \cos{\left[ \frac{2 \pi}{T} (s-1) \right]}\right) } \ ,
\end{aligned}
\end{equation}
where $R = t-t^\prime$ is the distance between the two lattice sites being considered. We can now approximate the above expression for $T \gg 1$ as follows:
\begin{equation}
\begin{aligned}
s_{t t^\prime} &\stackrel{T \gg 1}{\approx} \int_0^{\infty} \frac{dz}{2} e^{-z \lambda_2} \int_0^{2 \pi} \frac{d\omega}{2 \pi} e^{-z \lambda_3 \cos \omega} \cos{\left[\omega R\right]}  \\
&= \int_0^{\infty} \frac{dz}{2} e^{-z \lambda_2} I_R(-\lambda_3 z) =\\
&= \frac{\left(-\frac{\lambda_3}{\left|\lambda_3 \right|}\right)^R}{2} \int_0^{\infty} dz \, e^{-z \lambda_2} I_R(\left|\lambda_3 \right| z) \stackrel{\lambda_2 - \left|\lambda_3 \right| \ll 1  }{\approx} \\
&\approx \frac{\left(-\text{sign}(\lambda_3)\right)^R}{2} \int_0^{\infty} dz \, \frac{e^{-z \lambda_2 + \left|\lambda_3 \right| z - \frac{R^2}{2 \left|\lambda_3 \right| z} }}{\sqrt{2 \pi \left|\lambda_3 \right| z}}  \\
& = \frac{\left(-\text{sign}\left(\lambda_3\right) \right)^R}{2 \left|\lambda_3 \right| \sqrt{2 \frac{\lambda_2}{\left|\lambda_3 \right|}-2}} e^{-\left|R \right| \sqrt{2 \frac{\lambda_2}{\left|\lambda_3 \right|}-2} } \ ,
\end{aligned}
\end{equation}
where  $I_n(x)$ is the modified Bessel function of the first kind. Let us briefly comment on the approximation made in the third step of the above expression (i.e., for $\lambda_2 - \left|\lambda_3 \right| \ll 1$). As it can be seen from the expressions for $s_0$ and $s_1$ in Eq.~\eqref{eq: m,s0,s1,stt'}, such approximation  corresponds to a regime of strong time correlations up to lag-one. The approximation effectively becomes useful only to compute time correlations at lag two or higher, i.e., to compute $s_{t t^\prime}$ for $t^\prime > t+1$, given that those at lower lags are known exactly. Therefore, $\lambda_2 - \left|\lambda_3 \right| \approx 1$ corresponds to a regime of low time correlations even at lags one and zero (it should be noted here that correlations of the type $\langle x_t x_{t^\prime} \rangle$ are not normalised to one when $t = t^\prime$, as is instead the case with the standard definition of autocorrelation). This, in turn, ensures that time correlations at higher lags will be low enough to make the error due to the above approximation negligible.

We can now plug the above result into Eq.~\eqref{eq: plefka exp sph} via Eq.~\eqref{eq: sph moments} in order to compute the approximate form of the free energy deriving from the partition function in Eq.~\eqref{eq: Z quartic}. After having computed such approximate form for $F$, we can calculate the Lagrange multipliers as usual, i.e., by solving the system of equations $ \langle \mathcal{O}_\ell(X) \rangle = \partial F / \partial \beta_\ell \; , \forall \, \ell $. Alternatively, one could truncate Eq.~\eqref{eq: boh} to the second order of the couplings $\lambda_3$ and $\lambda_4$, find an approximate form $Z_{p}$ of $Z$ and then maximize the approximate likelihood $e^{-H(\overline{X})}/Z_{p}$.

\begin{figure*} 
  \centering
  \includegraphics[width=1\linewidth]{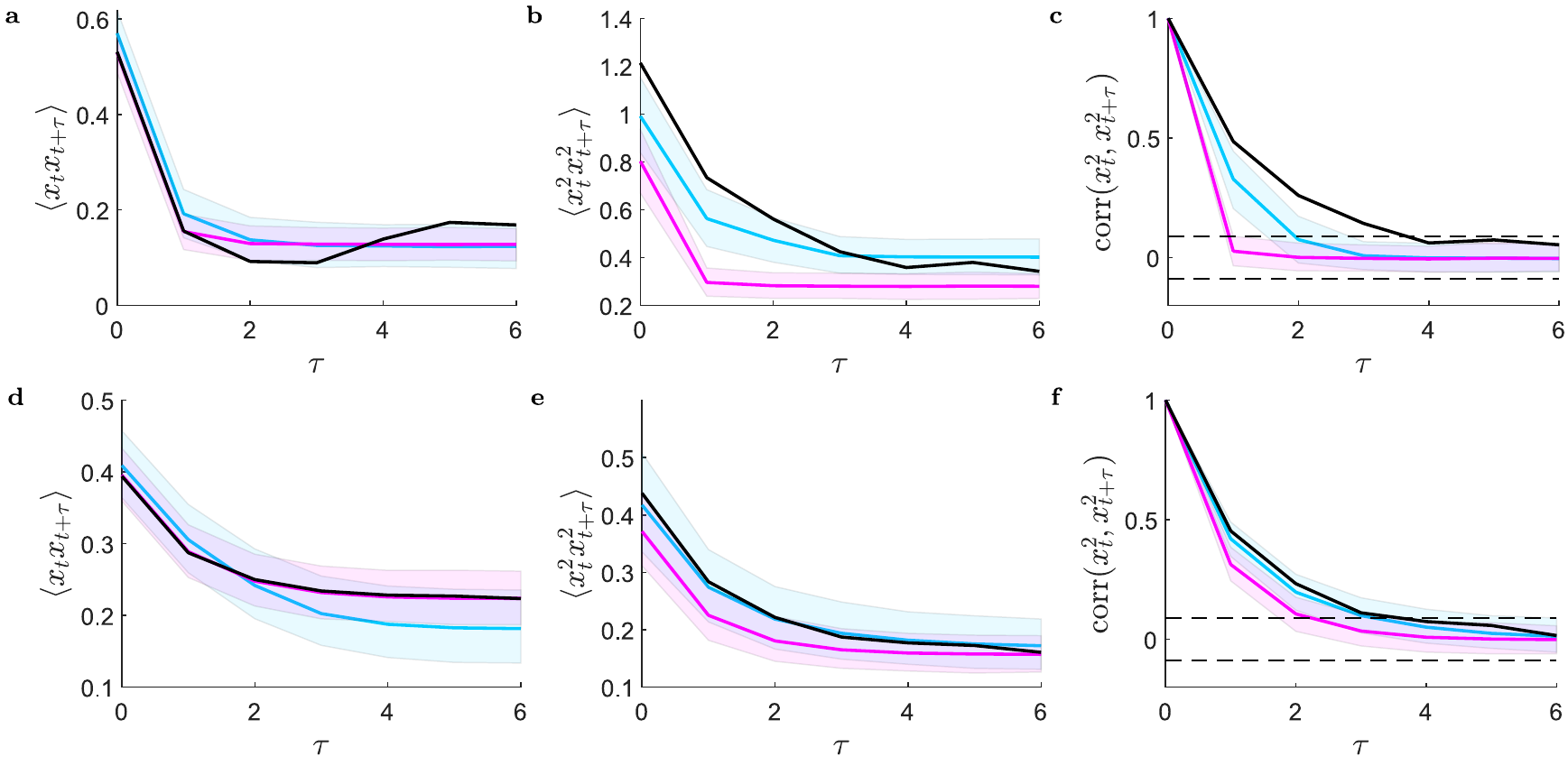}
  \caption{\textbf{Ability of the different models to match specified constraints.} Comparisons between empirical time correlations and the corresponding quantities as measured in the ensembles defined by Eq.~\eqref{eq: Z spherical} (purple/dark grey) and Eq.~\eqref{eq: Z quartic} (blue/light grey). Panels ({\bf a}) and ({\bf d}) refer to $\langle x_t x_{t+\tau} \rangle$, panels ({\bf b}) and ({\bf e}) to $\langle x_t^2 x_{t+\tau}^2 \rangle$, and panels ({\bf c}) and ({\bf f}) to $(\langle x_t^2 x_{t+\tau}^2 \rangle - \langle x_t^2 \rangle \langle x_{t+\tau}^2 \rangle )/\langle x_t^4 \rangle$. In the three upper panels the empirical correlations (black solid lines) are computed from one instance of the autoregressive model $Y_{t+1} = \xi_t^{(-1.5,1.5)} Y_t + \xi_t^{(-0.2,0.8)}$, whereas in the three lower panels correlations are computed from the model $Y_{t+1} = \xi_t^{(-0.375,1.125)} Y_t + \xi_t^{(-0.2,0.8)}$. In panels ({\bf c}) and ({\bf f}) horizontal dashed lines denote the $95\%$ confidence level interval for the autocorrelation of white noise.}
  \label{fig: quartic ens}
\end{figure*}

In Figure~\ref{fig: quartic ens} we show the ability of the ensemble introduced in this Section to match the imposed constraints with respect to its unperturbed counterpart. We do so using two autoregressive models with markedly distinct correlation features. The first model ($Y_{t+1} = \xi_t^{(-1.5,1.5)} Y_t + \xi_t^{(-0.2,0.8)}$, panels ({\bf a}-{\bf c})) is designed to produce time series that - on average - have non-zero correlations only between second or higher order moments. This represents an ``adversarial'' example, in the sense that the only correlations present in the process cannot be captured by the unperturbed model of Eq.~\eqref{eq: Z spherical}. This would suggest the need to use a ``stronger'' perturbation than the second order one in order to substantially improve the model's ability to capture the correlations of the process. However, panels ({\bf b}) and ({\bf c}) show that even stopping the perturbation expansion at the second order gives a sizeable improvement. The second model ($Y_{t+1} = \xi_t^{(-0.375,1.125)} Y_t + \xi_t^{(-0.2,0.8)}$, panels ({\bf d}-{\bf f})) is instead designed to produce time series with time correlations between first moments as well. This represents a scenario where the unperturbed model captures by design correlations between first moments, which translates into a partial ability to capture higher order correlations. These are then fully captured by the full model (see panels ({\bf e}) and ({\bf f})).

\floatsetup[figure]{style=plain,subcapbesideposition=top}

In order to quantitatively assess the improvement of the proposed perturbative solution with respect to the unperturbed ensemble, we report the results from a few simple prediction exercises in Table~\ref{tab: pred exercise}. Namely, we seek to predict the mean and the $10\%$ and $90\%$ quantiles of the data generating process one lag ahead. We compare such quantities against those computed from both the unperturbed and full ensemble, and we quantify the agreement with two widely adopted metrics of accuracy, namely the root mean square error (RMSE) and the $R^2$. As it can be seen, in all cases switching from the unperturbed ensemble to the one in Eq.~\eqref{eq: Z quartic} systematically provides a measurable improvement, regardless of the specific model considered. Notably, the biggest relative improvement occurs for the model that we identified as an ``adversarial'' example. This is because, as mentioned above, in that case the unperturbed model cannot capture by design the relevant time correlations in the data generating process.

\begin{table}[]
\begin{tabular}{c|cccccc|}
\cline{2-7}
\multirow{2}{*}{} &
  \multicolumn{2}{c|}{$\mathbf{M_1}$} &
  \multicolumn{2}{c|}{$\mathbf{M_2}$} &
  \multicolumn{2}{c|}{$\mathbf{M_3}$} \\ \cline{2-7} 
 &
  \multicolumn{1}{c|}{$\mathbf{RMSE}$} &
  \multicolumn{1}{c|}{$R^2$} &
  \multicolumn{1}{c|}{$\mathbf{RMSE}$} &
  \multicolumn{1}{c|}{$R^2$} &
  \multicolumn{1}{c|}{$\mathbf{RMSE}$} &
  $R^2$ \\ \hline
\multicolumn{1}{|c|}{$\mathbf{\overline{x}_{H_1}}$} & 0.0267 & 0.995 & 0.125  & 0.923 & 0.204 & 0.938 \\ \cline{1-1}
\multicolumn{1}{|c|}{$\mathbf{\overline{x}_{H_2}}$} & 0.0155 & 0.998 & 0.0818 & 0.969 & 0.176 & 0.954 \\ \cline{1-1}
\multicolumn{1}{|c|}{$\mathbf{q^{0.9}_{H_1}}$}      & 0.0900 & 0.943 & 0.236  & 0.847 & 0.277 & 0.866 \\ \cline{1-1}
\multicolumn{1}{|c|}{$\mathbf{q^{0.9}_{H_2}}$}      & 0.0511 & 0.985 & 0.122  & 0.957 & 0.232 & 0.910 \\ \cline{1-1}
\multicolumn{1}{|c|}{$\mathbf{q^{0.1}_{H_1}}$}      & 0.0825 & 0.960 & 0.206  & 0.887 & 0.170 & 0.970 \\ \cline{1-1}
\multicolumn{1}{|c|}{$\mathbf{q^{0.1}_{H_2}}$}      & 0.0495 & 0.975 & 0.188  & 0.906 & 0.151 & 0.976 \\ \hline
\end{tabular}
\caption{Accuracy of one lag ahead predictions of the mean and $10\%$ and $90\%$ quantiles of the three data generating processes used so far. These are denoted respectively as $M_1$ ($Y_{t+1} = \xi_t^{(-0.375,1.125)} Y_t + \xi_t^{(-0.2,0.8)}$), $M_2$ ($Y_{t+1} = \xi_t^{(-1.5,1.5)} Y_t + \xi_t^{(-0.2,0.8)}$), and $M_3$ ($Y_{t+1} = \xi_t^{(0,1.5)} Y_t + \xi_t^{(-0.3,0.7)}$). The means and quantiles are denoted as $\mathbf{\overline{x}}$ and $\mathbf{\overline{q}}$. $H_1$ and $H_2$ denote, respectively, predictions obtained by means of the unperturbed ensemble of Eq.~\eqref{eq: Z spherical} and the full ensemble of Eq.~\eqref{eq: Z quartic}.}
\label{tab: pred exercise}
\end{table}

\section{Conclusions}

In this paper we have shown how we can apply tools from classical Statistical Mechanics to time series analysis by simply mapping the time dimension of a time-evolving system onto the spatial dimension of a lattice. This allows to design ensembles that preserve - on average - some desired constraints on the temporal structure of the data under consideration by imposing constraints on such a lattice.

In particular, we have shown how a constraint on the lag-one autocorrelation corresponds to a well known ensemble in Statistical Mechanics, namely the Spherical Model of a (anti/)ferromagnet. Moreover, we have shown how the Spherical Model can be used as a basis to handle higher order temporal correlations by using a perturbation theory approach. We have also shown how the inferred Lagrange multipliers of the ensembles can be updated in ``real time'' as new data from the system of interest are collected, which in turn allows to obtain information about the possible evolution of the underlying data generating process. 


The framework presented here, coupled with tools commonly used to tackle inverse Ising problems such as Pseudo-Likelihood methods, can be adapted in order to handle multiple time series and their correlations. Moreover, the accuracy of the single time series case presented here can be improved by considering higher perturbation orders or by considering different Hamiltonians. In particular, it would be interesting to extend the approach proposed here to Hamiltonians whose Lagrange multipliers are drawn from parametric distributions (similarly to couplings in spin-glass systems), which would provide an alternative - and possibly even more flexible - method to fit ensembles to some desired constraints. We hope to see some of these topics pursued in the near future.

\begin{center}
\textbf{Acknowledgments}
\end{center}
G.L. acknowledges support from an EPSRC Early Career Fellowship (Grant No. EP/N006062/1).

\newpage

\bibliography{apssamp}
\clearpage
\appendix
\onecolumngrid
\section{Polynomial functions used in Eq.~\eqref{eq: sph moments}}
\label{app:formulas}
\begin{equation*}
    \begin{aligned}
        &\langle x_t^4 x_{t^\prime}^2 x_{t^\prime+1}^2 \rangle_0 =&&  m^8+12 m^2 s_0^3+2 m^4 \left(4 \left(m^2+2 s_1\right) s_{t t'}+s_1 \left(2 m^2+s_1\right)+6 s_{t t'}^2\right) \\ & \; && +8 m^2 s_{(t+1) t'} \left(m^4+6 \left(m^2+s_1\right) s_{t t'} +2 m^2 s_1+6 s_{t t'}^2\right)+12 s_{(t+1) t'}^2 \left(m^4+2 s_{t t'} \left(2 m^2+s_{t t'}\right)\right)\\& \; && +4 s_0 \left[2 m^2 \left(m^2+s_{t t'}+s_{(t+1) t'}\right) \left(m^2+3 s_{t t'}+3 s_{(t+1) t'}\right)+3 m^2 s_1^2 \right. \\& \; && \left. + 6 s_1 \left(m^4+2 m^2 s_{t t'}+2 s_{(t+1) t'} \left(m^2+s_{t t'}\right)\right)\right]\\& \; && +2 s_0^2 \left(8 m^4+6 \left(2 m^2 s_{t t'}+2 m^2 s_{(t+1) t'}+s_{t t'}^2+s_{(t+1) t'}^2\right) +6 m^2 s_1+3 s_1^2\right)+3 s_0^4\\
        &\langle x_{t}^2 x_{t+1}^2 x_{t^\prime}^2 x_{t^\prime+1}^2 \rangle_0 =&& m^8+4 s_0^3 m^2+16 s_1^3 m^2+8 s_1 \left[\left(m^2+2 s_{(t+1) t'}\right) \left(m^2+2 s_{t \left(t'+1\right)}\right) \right. \\& \; && \left.+\left(2 m^2+s_{(t+1) t'}+s_{t \left(t'+1\right)}\right) s_{(t+1) \left(t'+1\right)}+s_{t t'} \left(2 m^2+s_{(t+1) t'}+s_{t \left(t'+1\right)}+4 s_{(t+1) \left(t'+1\right)}\right)\right] m^2 \\& \; && +s_0^4+4 s_1^4+2 s_0^2 \left[3 m^4+4 s_1 m^2+2 s_{t t'} m^2+2 s_{(t+1) t'} m^2+2 s_{t \left(t'+1\right)} m^2+2 s_{(t+1) \left(t'+1\right)} m^2 \right. \\& \; && \left. +2 s_1^2+s_{t t'}^2+s_{(t+1) t'}^2+s_{t \left(t'+1\right)}^2+s_{(t+1) \left(t'+1\right)}^2\right]+4 s_1^2 \left[5 m^4+4 \left(s_{t \left(t'+1\right)}+s_{(t+1) \left(t'+1\right)}\right) m^2 \right. \\& \; && \left. +4 s_{(t+1) t'} \left(m^2+s_{t \left(t'+1\right)}\right)+4 s_{t t'} \left(m^2+s_{(t+1) \left(t'+1\right)}\right)\right] \\& \; && +2 \left[ \left(s_{(t+1) \left(t'+1\right)}^2+2 \left(m^2+2 s_{t \left(t'+1\right)}\right) s_{(t+1) \left(t'+1\right)}+s_{t \left(t'+1\right)} \left(2 m^2+s_{t \left(t'+1\right)}\right)\right) m^4 \right. \\& \; && \left. +2 s_{(t+1) t'} \left(m^4+2 s_{t \left(t'+1\right)} \left(2 m^2+s_{t \left(t'+1\right)}\right)+2 \left(m^2+2 s_{t \left(t'+1\right)}\right)  s_{(t+1) \left(t'+1\right)}\right) m^2 \right. \\& \; && \left. +s_{(t+1) t'}^2 \left(m^4+2 s_{t \left(t'+1\right)} \left(2 m^2+s_{t \left(t'+1\right)}\right)\right) +2 s_{t t'} \left(2 s_{(t+1) \left(t'+1\right)}^2 m^2 \right. \right. \\& \; && \left. \left. +\left(m^2+2 s_{(t+1) t'}\right) \left(m^2+2 s_{t \left(t'+1\right)}\right) m^2+4 \left(m^2+s_{(t+1) t'}\right) \left(m^2+s_{t \left(t'+1\right)}\right) s_{(t+1) \left(t'+1\right)}\right) \right. \\& \; && \left. +s_{t t'}^2 \left(m^4+2 s_{(t+1) \left(t'+1\right)} \left(2 m^2+s_{(t+1) \left(t'+1\right)}\right)\right)\right]+4 s_0 \left[2 s_1^2 m^2 \right. \\& \; && \left. +\left(s_{t t'}^2+2 \left(m^2+s_{(t+1) t'}+s_{t \left(t'+1\right)}\right) s_{t t'}+s_{(t+1) t'}^2 \right. \right. \\& \; && \left. \left. +\left(m^2+s_{t \left(t'+1\right)}+s_{(t+1) \left(t'+1\right)}\right){}^2+2 s_{(t+1) t'} \left(m^2+s_{(t+1) \left(t'+1\right)}\right)\right) m^2 \right. \\& \; && \left. +2 s_1 \left(2 \left(m^2+s_{(t+1) t'}+s_{t \left(t'+1\right)}\right) m^2 \right. \right. \\& \; && \left. \left. +s_{t t'} \left(2 m^2+s_{(t+1) t'}+s_{t \left(t'+1\right)}\right) +\left(2 m^2+s_{(t+1) t'}+s_{t \left(t'+1\right)}\right) s_{(t+1) \left(t'+1\right)}\right)\right]
    \end{aligned}
\end{equation*}
\end{document}